\documentclass[fleqn,twoside]{article}
\usepackage{espcrc2}

\usepackage{epsf}

\usepackage{graphicx}
\usepackage{dcolumn}
\usepackage{bm}

\begin{document}

\title{Charmonium Spectrum from Quenched QCD with Overlap Fermions}
\author{S. Tamhankar\address[UK]{Department of Physics and Astronomy, 
University of Kentucky, Lexington, KY 40506, USA},
A. Alexandru\addressmark[UK], Y. Chen\address{Institute of High Energy
Physics, 
Chinese Academy of Sciences, Beijing 100039, P. R. China}, S. J.
Dong\addressmark[UK], T. Draper\addressmark[UK], I.
Horv\'{a}th\addressmark[UK],\\F. X. Lee\address{George Washington
University, Washington, DC 20052 USA\\ 
        Jefferson Lab, 12000 Jefferson Avenue, Newport News, VA 23606, 
USA}, K. F. Liu\addressmark[UK], N. Mathur\addressmark[UK], J. B.
Zhang\address{CSSM and Department of Physics and Mathematical Physics,
University of Adelaide, Adelaide, SA 5005, Australia}}
%
\begin{abstract}
\vspace{-0.3cm}
We present preliminary results using overlap
fermions for the charmonium spectrum, 
in particular for hyperfine splitting. Simulations are
performed on $16^3 \times 72$ lattices, with Wilson gauge action at
$\beta=6.3345$. Depending on how the scale is set, we obtain 104(5)~MeV
(using $1\bar{P}-1\bar{S}$) or 88(4)~MeV
(using $r_0$=0.5~fm) for the hyperfine splitting.
%
\vspace{-0.8cm}
\end{abstract}

\maketitle
\section{Introduction}
\vspace{-0.2cm}
Overlap fermions have
the following desirable features:
\begin{itemize}
\vspace{-0.2cm}
\item{Exact chiral symmetry on the lattice.}
\vspace{-0.2cm}
\item{No additive quark mass renormalization.}
\vspace{-0.2cm}
\item{No flavor symmetry breaking.}
\vspace{-0.2cm}
\item{No $\mathcal{O}(a)$ error.}
\vspace{-0.2cm}
\item{The $\mathcal{O}(m^2a^2)$ and $\mathcal{O}(\Lambda_{\rm QCD}ma^2)$ errors are also small, from dispersion relation
and renormalization constants.}
\vspace{-0.2cm}
\end{itemize}

The first two features are especially significant for light quarks.
Many exciting results at low quark masses have been reported using
overlap fermions. Here we want to make the point that using overlap
fermions can
also alleviate some problems with simulating heavy quarks. The last
feature, demonstrated in~\cite{liu02},
is an unexpected bonus in this regard. The key observation is that the
discretization errors are only about 5\% all the way upto $ma \approx 0.5$.

We make our case using hyperfine splitting in the 
charmonium system.
It is known that with staggered quarks, there is an ambiguity about
Nambu-Goldstone (NG) and non-NG mode for the $\eta_c$,
resulting
in widely different estimates of hyperfine splitting --
51(6)~MeV (non-NG) and 404(4)~MeV (NG)~\cite{aoki}. NRQCD converges only
slowly for charm~\cite{NR}. Including
$\mathcal{O}(v^6)$ terms
changed the result from 96(2)~MeV to 55(5)~MeV. Wilson fermions have
$\mathcal{O}(a)$
errors. Hyperfine splitting is very
sensitive to the coefficient of the correction term, $c_{SW}$.
There are many studies \cite{wilson} using Wilson type valence quarks, including some with
non-perturbative $c_{SW}$. The quenched clover estimate of hyperfine
splitting has stabilized
around 80 MeV. Result from a 2+1 dynamical simulation using tree-level $c_{SW}$ still
falls short of the experimental value by about 20\%~\cite{dyn}.

Although costly, overlap fermions offer the best solution --- they do not have flavor
symmetry breaking, they guarantee that the $\mathcal{O}(a)$ error is absent {\it without}
any tuning and they are relativistic. Furthermore, $\mathcal{O}(a^2)$ errors
appear to be small~\cite{liu02}. 
\vspace{-0.3cm}
\section{Simulation Details}
\vspace{-0.3cm}
\begin{figure}
\vspace{-0.5cm}
\includegraphics[width=8cm]{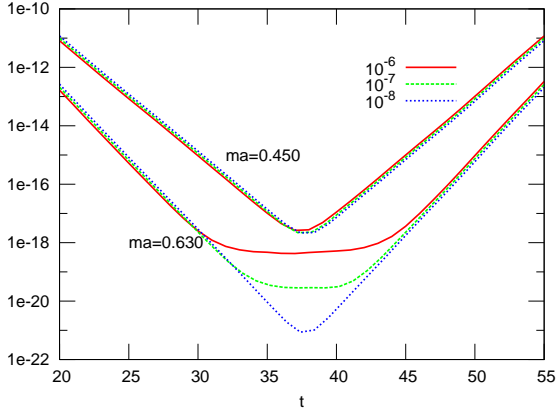}
\vspace{-1.5cm}
\caption{Effect of inner loop precision on pseudoscalar propagators for
heavy
quarks. We study output of one spin and one color for a single
configuration for this illustration. Curves are slightly shifted for
clarity.  \label{fig-inner}}
\vspace{-0.5cm}
\end{figure}
\begin{figure}[h]
\vspace{-0.8cm}
\includegraphics[width=8cm]{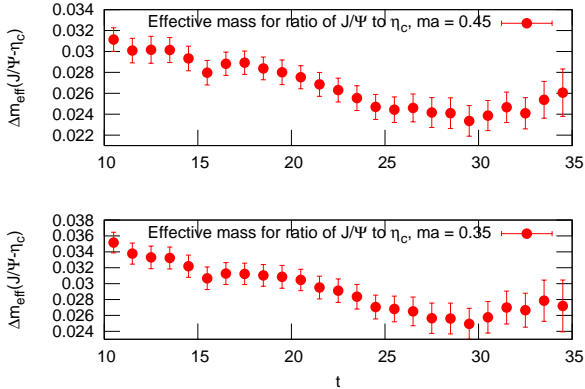}
\vspace{-1.5cm}
\caption{Effective hyperfine splitting ($m_{J/\psi}-m_{\eta_c}$)
from the ratio of the vector to pseudoscalar
correlators for two
values of $ma$ which bracket the charm mass. \label{fig-eff}}
\vspace{-1.0cm}
\end{figure}
\vspace{0.2cm}
Our simulations are performed on $16^3 \times 72$ isotropic lattices. We
present results on 100 configurations. The Wilson gauge action is used at
$\beta$ = 6.3345. For quenched configurations, the scales set from
different physical quantities can differ considerably. Hyperfine splitting is
very sensitive to the scale, so we report results for scale set using
two different quantities: $r_0$ and the ($1\bar{P}-1\bar{S}$) mass splitting
in charmonium. From ($1\bar{P}-1\bar{S}$)  we obtain $a=0.0501$~fm.
Using $r_0=0.5$~fm the scale is $a=0.0560$~fm~\cite{sommer}. 
We use a multimass inverter to obtain propagators for  26 masses 
ranging from 0.020--0.85 in lattice
units. The bare masses correspond to 70 MeV -- 3 GeV. The charm mass in
lattice units is 0.365(1) for the ($1\bar{P}-1\bar{S}$) scale, and 0.431(1)
for $r_0$ scale. Near these values of $ma$ our discretization
errors are estimated to be under 5\%.

For heavy quarks, mesonic two-point functions
fall through many orders of magnitude, and become very small at the
center of the lattice. An imprecise quark propagator results in
drastically different values for these small numbers. Our observation is
that a good inner loop precision is crucial, as shown in 
Fig.~\ref{fig-inner}. To confirm that our outer loop precision is
adequate, we increased it by an order of magnitude. This lead to less
than a percent deviation in the (single configuration, single spin-color)
pseudoscalar correlator.
For our production runs, we
choose an inner loop precision of $10^{-8}$ and outer loop precision of
$10^{-5}$, with eigenvalue projection precision of $10^{-10}$. At charm
mass, the residual is about $2 \times 10^{-9}$.
\vspace{-0.2cm}
%
                                                                                
\begin{figure}[t]
\vspace{-0.5cm}
\includegraphics[width=8cm]{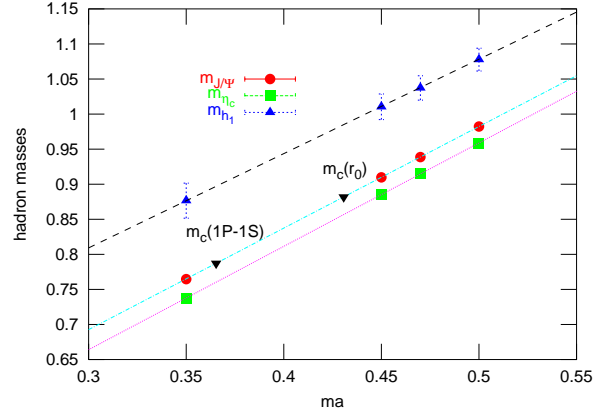}
\vspace{-1.5cm}
\caption{We fit the meson masses linearly in quark mass.  Fits are shown
for
$\eta_c,J/\psi$ and $h_c$ masses. \label{fig-inter}}
\vspace{-1.0cm}
\end{figure}

\section{Analysis}
\begin{figure}[h]
\includegraphics[width=8cm]{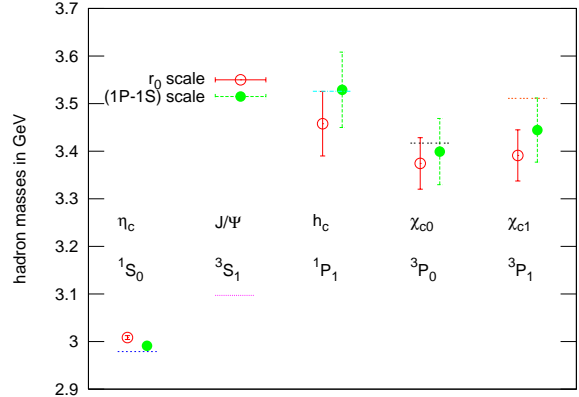}
\vspace{-1.3cm}
\caption{Charmonium spectrum in physical units. Results from both $r_0$
and
$1\bar{P}-1\bar{S}$ scales are shown. Note, for the latter scale, a
linear combination of
$h_c$ and $\eta_c$ masses, along with the $J/\psi$ mass, is used for
input.\label{fig-spectrum}}
\vspace{-0.8cm}
\end{figure}

\begin{figure}[h]
\includegraphics[width=8cm]{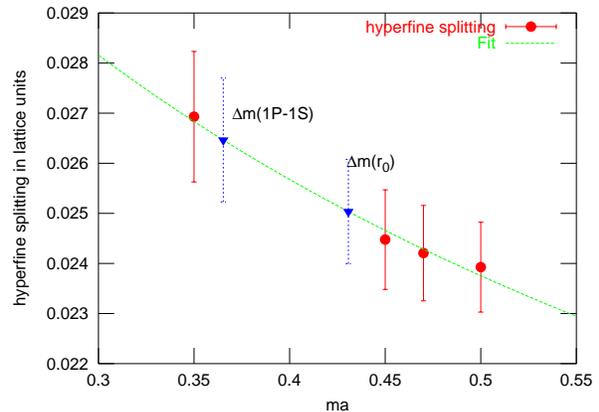}
\vspace{-1.5cm}
\caption{Hyperfine splitting as a function of quark mass, with
interpolation shown at $m_ca$.\label{fig-hyp}}
\vspace{-0.8cm}
\end{figure}

\vspace{-0.2cm}
Fig.~\ref{fig-eff} shows the effective hyperfine splitting from the
ratio of vector to pseudoscalar correlator.
We use two ways to set the scale -- 
from the ($1\bar{P}-1\bar{S}$)
splitting in the charmonium system and from $r_0$ (using 0.5~fm).
The experimental $m_{J/\psi}$ is used to set $m_c$ (in lattice
units). This is straight-forward when the 
scale from $r_0$ is used.
The singlet $P$ mass $m_{h_c}$ is used for $\bar{P}$,
and $(3 m_{J/\psi} + m_{\eta_c})/4$ for $\bar{S}$ mass.
The spin averaged ($1\bar{P}-1\bar{S}$) splitting is used because
it is expected to be
insensitive to lattice artifacts. In this case, the determination of $a$ and
$m_ca$ is entangled. The procedure we follow to disentangle these is as
follows.
As shown in Fig.~\ref{fig-inter} all hadron masses in lattice units are fitted to a straight line,
$m_ha = A_h.ma + B_h$.
Lattice spacing $a$ and bare charm quark mass
$m_ca$ are
two
unknowns; $m_{J/\psi}$ and
$m(1\bar{P}-1\bar{S})$ in physical units are the
two inputs. We solve for $a$ and $m_ca$ to obtain the values quoted
earlier. If we were to use this $a$ to get the Sommer scale, we would obtain
$r_0$ = 0.45 fm. Fig.~\ref{fig-spectrum} shows the charmonium spectrum
in physical units. Interpolation for the hyperfine splitting is shown in
Fig.~\ref{fig-hyp}.
The fit
form used is $(m_{J/\psi}-m_{\eta_c})a = A/\sqrt{ma}+B/ma$. The interpolation for $m_c$
for the two scales is shown. Note, hyperfine splitting is higher for ($1\bar{P}-1\bar{S}$)
scale in lattice units and also $a$ is smaller, so the result in MeV is considerably higher
for that case.
Finally we summarize the results in the table below. All masses are in
GeV.
\vspace{-0.6cm}
\begin{table}[h]
\begin{center}
\begin{tabular}{|c|c|c|c|}
\hline
\vspace*{-0.3cm}
&&&\\
&$a(r_0)$ &$a(1\bar{P}-1\bar{S})$ & Expt \\
\vspace*{-0.35cm}
&&&\\
\hline
\vspace*{-0.35cm}
&&&\\
$\eta_c$&3.008(4)&2.991(6)&2.980\\
$J/\psi$&---&---&3.097\\
$J/\psi-\eta_c$&0.088(4)&0.104(5)&0.117\\
$h_c$&3.46(7)\ \ &3.53(8)\ \ &3.526\ \\
$\chi_{c0}$&3.37(5)\ \ &3.40(7)\ \ &3.41\ \ \\
$\chi_{c1}$&3.39(5)\ \ &3.44(7)\ \ &3.511\ \\
\hline
\end{tabular}
\vspace{0.3cm}
\caption{Charmonium spectrum (GeV) \label{table_spectrum}}
\end{center}
\end{table}
\vspace{-1.7cm}
\section{Summary}
\vspace{-0.2cm}
We have presented the first study of the charmonium spectrum using overlap
fermions. We get a better agreement with the experimental spectrum using
$1\bar{P}-1\bar{S}$ scale rather than the $r_0$ scale.
Our value for hyperfine splitting is 104(5)~MeV and 88(4)~MeV
using $1\bar{P}-1\bar{S}$ and $r_0$ scale respectively.
This is considerably higher than the quenched clover
results. We make two approximations: quenching and excluding OZI
suppressed
diagrams.
At the charm mass the singlet contribution appears to be
small~\cite{disconn}, though lattice calculations with smaller
statistical and systematic errors are needed to settle this issue.
If the unquenched clover simulations are any indication,
then it is possible that dynamical overlap with only connected
insertion will reproduce the experimental result.
                                                                                
This work is supported in part by U.S. Department of Energy
under grants DE-FG05-84ER40154 and DE-FG02-95ER40907. The computing
resources at
NERSC (operated by DOE under DE-AC03-76SF00098) are also acknowledged.
Y. Chen and S. J. Dong are partly supported by NSFC (\#10235040 and
\#10075051 )

\begin{thebibliography}{9}
\bibitem{liu02} K.~F.~Liu and S.~J.~Dong, hep-lat/0206002.
\bibitem{aoki} S.~Aoki {\it et al.} Nucl.\ Phys.\ Proc.\ Suppl.\ {\bf
42}, 303 (1995).
\bibitem{NR}
C.~T.~H. Davies {\em et~al.}, Phys. Rev. {\bf D52}, 6519 (1995), 
H.~D. Trottier, Phys. Rev. {\bf D55}, 6844 (1997). 
\bibitem{wilson}
An incomplete list: C.~R.~Allton {\em et~al.}, Phys. Lett. {\bf B292},
408 (1992), M.~Okamoto {\it et al.} Phys.\ Rev.\ D {\bf 65}, 094508
(2002), S.~Choe {\it et al.}  JHEP {\bf 0308}, 022 (2003).
\bibitem{dyn} M.~di Pierro
{\it et al.}, Nucl.\ Phys.\ Proc.\ Suppl.\  {\bf 129}, 340 (2004).
\bibitem{sommer} S.~Necco and R.~Sommer Nucl.\ Phys.\ {\bf B622}, 328
(2002).
\bibitem{disconn}
C. McNeile and C. Michael, hep-lat/0402012,
P.~de Forcrand {\it et al.} hep-lat/0404016.
\end{thebibliography}
\end{document}